\documentclass[prb,preprint]{revtex4} 
\usepackage{amsmath}

\usepackage{graphicx}
\usepackage{mathrsfs}

\preprint{\tiny The following article has been accepted by 
the American Journal of Physics. After it is published, it will be 
found at http://scitation.aip.org/ajp.}

\begin{document}

\title{Dynamics of a double pendulum with distributed mass}

\author{M. Z. Rafat}
\email{rafat@physics.usyd.edu.au}
\affiliation{School of Physics, University of Sydney, NSW 2006, 
Australia}
\author{M. S. Wheatland}
\email{m.wheatland@physics.usyd.edu.au}
\affiliation{School of Physics, University of Sydney, NSW 2006, 
Australia}
\author{T. R. Bedding}
\email{t.bedding@physics.usyd.edu.au}
\affiliation{School of Physics, University of Sydney, NSW 2006, 
Australia}

\begin{abstract}
We investigate a variation of the simple double pendulum in which 
the two point masses are replaced by square plates. The double square
pendulum exhibits richer
behavior than the simple double pendulum and provides a convenient 
demonstration of nonlinear dynamics and chaos. It is also an example 
of an asymmetric compound double pendulum, which has not been 
studied in detail. We obtain the equilibrium
configurations and normal modes of oscillation and derive the 
equations of
motion, which are solved numerically to produce Poincar\'{e} sections. 
We show how the behavior varies from regular motion at low energies, 
to chaos at intermediate energies, and back to regular motion at high 
energies. We also show that the onset of chaos occurs at a 
significantly lower energy than for the simple double pendulum.
\end{abstract}

\maketitle

\section{Introduction\label{sec:intro}}

The simple double pendulum consisting of two point masses 
attached by massless rods and free to rotate in a plane is one of the 
simplest dynamical systems to exhibit 
chaos.\cite{richter84,korsch99,stachowiak06} 
It is also a prototypical system for demonstrating the Lagrangian and 
Hamiltonian approaches to dynamics and the machinery of nonlinear 
dynamics.\cite{landau76,goldstein80,kibble05,gregory06} Variants of
the simple double pendulum have been considered, including an 
asymmetrical version,\cite{newton89} and a configuration in which
the inner mass is displaced along the rod.\cite{levien93}
The compound (distributed-mass) double pendulum 
is a generalization 
that is easier to implement as a demonstration. For example, the 
double bar
pendulum in which the point masses are replaced by slender bars has
been the subject of a number of 
studies,\cite{shinbrot92,zhou96,ohlhoff00}
and a version of the double square pendulum is available 
commercially.\cite{chaoticpendulums} The dynamics of the general 
symmetrical compound double pendulum has also been investigated, 
including a proof that it is a chaotic system.\cite{dullin94}

\begin{figure}[!ht]
\begin{center}
\scalebox{0.75}{\includegraphics{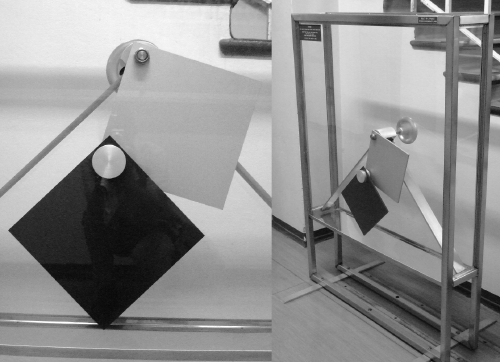}}
\caption{\label{fig:pic}The double square pendulum in the School of 
Physics at the University of Sydney. Left: close up of the 
two plates. Right: the pendulum and its enclosure.}
\end{center}
\end{figure}

The School of Physics at the University of Sydney has a large-scale
variation of the compound planar double pendulum (see
Fig.~\ref{fig:pic}). The double square pendulum consists of two square 
metal plates connected together by two axles. It is set into motion 
by rotating a wheel at the back, which is attached to the axle on the
inner plate. The axles are housed in low friction bearings,
so that when the wheel is turned and released, the plates continue in 
a complex motion that lasts several minutes.\cite{wheatland08a}
The demonstration is housed inside a
large glass and metal enclosure measuring about 120\,cm by 120\,cm by
25\,cm, and the square metal plates are approximately 28\,cm on a side.
The double square pendulum is located in the main corridor of the 
building and attracts considerable attention from passing students. 
The School also has a smaller, bench-mounted version of the double 
square pendulum, suitable for classroom demonstrations.

The double square pendulum exhibits diverse and at
times unpredictable behavior. For small pushes on the driving wheel, the
system oscillates back and forth about the equilibrium position shown in
Fig.~\ref{fig:pic}. If the driving wheel is rotated very rapidly, the
inner plate spins rapidly and throws the other plate outward, and the
motion is again fairly regular. For intermediate rates of rotation of the
driving wheel the system exhibits unpredictable motions.

This general behavior is similar to that of the simple and compound planar
double pendula.\cite{korsch99, ohlhoff00} However, the double square
pendulum described here differs from previously studied systems. It is a
compound pendulum, but the center of mass of the inner plate does not
lie along the line joining the two axles. In this regard it is
closest to the asymmetrical double pendulum studied in
Ref.~\onlinecite{newton89}, although that pendulum does not have distributed 
mass. The dynamics
of the double square pendulum is interesting both from the viewpoint of 
elucidating the
physics of a classroom physics demonstration and as a pedagogical exercise 
illustrating the application 
of dynamical theory.

In this paper we investigate an idealized model of the double square
pendulum. We first obtain the equations of motion using the Lagrangian
formalism (Sec.~\ref{sec:eqns}). General statements are then made
about the
basic motion of the double pendulum: the energy ranges for different types
of motion of the pendulum are identified, and the behavior at
low energy (Sec.~\ref{sec:bchar:normal}) and at high
energy (Sec.~\ref{sec:bchar:high}) is described. The equations of 
motion are
expressed in dimensionless form in Sec.~\ref{sec:num} and solved numerically
to produce Poincar\'{e} sections of the phase space of the pendulum, which
are constructed at increasing values of the total energy in
Sec.~\ref{sec:poinc}. Section~\ref{sec:device}
includes a brief qualitative comparison with the real double square pendulum.

\section{Equations of motion \label{sec:eqns}}

\begin{figure}[!ht]
\begin{center}
\scalebox{0.5}{\includegraphics{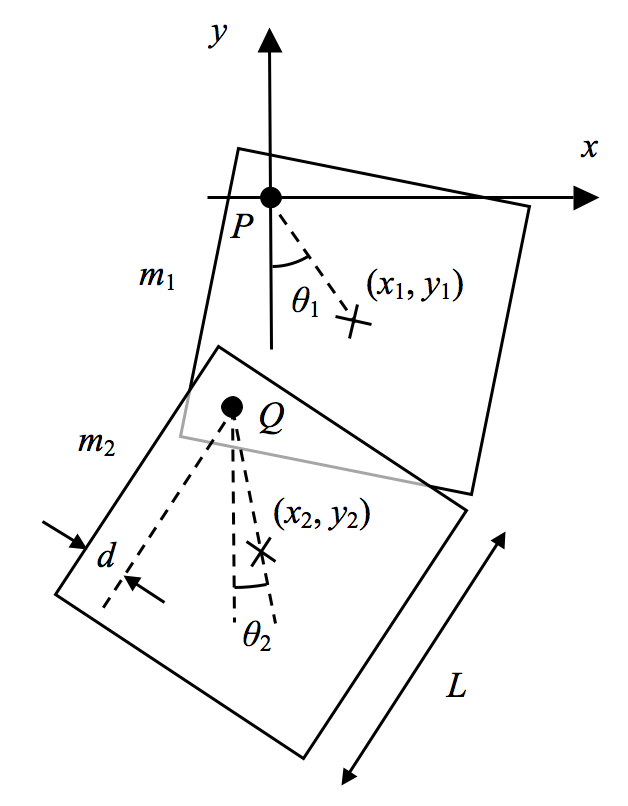}}
\caption{\label{fig:model}A model of the double square pendulum.}
\end{center}
\end{figure}

Consider a double pendulum comprising two square plates with side length $
L $ and masses $m_1$ and $m_2$ (see Fig.~\ref{fig:model}). The
inner plate rotates about a fixed axle at $P$ and the outer plate rotates
about an axle fixed to the inner plate at $Q$. We neglect the effect of
friction at the axles. The plates are assumed to have uniform mass
densities and the axles are assumed to be massless, so that the center of 
mass of each
plate is at its center. A coordinate system with origin at $P$ is defined
as shown and the center of mass of the inner and outer plates are located 
at positions $(x_1, y_1)$ and $(x_2, y_2)$, respectively. The center of 
mass of the plates subtend angles
$\theta_1$ and $\theta_2$ with respect to the direction of the negative $y$
axis.

The double square pendulum shown in Fig.~\ref{fig:pic} has relatively large 
axles, and in
particular the central axle has a wheel attached to it. The presence of
massive axles changes the locations of the center of mass of each
plate. However, the simple model considered here is expected to capture 
the essential dynamics of the real double square pendulum.

The equations of motion of the model pendulum may be derived using
Lagrangian dynamics.\cite{landau76,goldstein80,kibble05,gregory06} 
The Lagrangian is $\mathcal{L} = T - V$,
where $T$ is the kinetic energy and $V$ is the potential energy of 
the pendulum. The kinetic energy is
the sum of the kinetic energies of 
the center of mass of the two plates, each of which has a linear and a 
rotational component:
\begin{equation}\label{eqn:T_gen}
T = \frac{1}{2} m_1 (\dot{x}_1^2 + \dot{y}_1^2) 
+ \frac{1}{2} I_1 \dot{\theta}_1^2 
+ \frac{1}{2} m_2 (\dot{x}_2^2 + \dot{y}_2^2) + \frac{1}{2} I_2 \dot{\theta}_2^2,
\end{equation}
where $I_1$ and $I_2$ are the moments of inertia about the center of 
mass, and are given by $I_i = \frac{1}{6} m_i L^2$ for $i=1,2$.
The potential energy is the sum of the potential energy of each plate, 
and is given by
\begin{equation}\label{eqn:V_gen}
V = m_1 g y_1 + m_2 g y_2 + V_0,
\end{equation}
where $V_0$ is a suitable reference potential.

If we express $y_1$ and $y_2$ in terms of 
$\theta_1 $ and $\theta_2$, we obtain
\begin{equation}\label{eqn:V+V0}
V = -2 m_2 k_1\left[\cos (\theta_1 - \alpha) + \sin\alpha\cos \theta_2 \right]
+V_0,
\end{equation}
where
\begin{equation}\label{eqn:alpha}
\alpha = \tan^{-1} \frac{m_2}{m_1 + m_2},
\end{equation}
and 
\begin{equation}
k_1 = \frac{\sqrt{2}g\ell }{4} \mathrm{cosec}\,\alpha,
\end{equation}
with $\ell=L-2d$. 

\begin{figure}[!ht]
\begin{center}
\scalebox{0.43}{\includegraphics{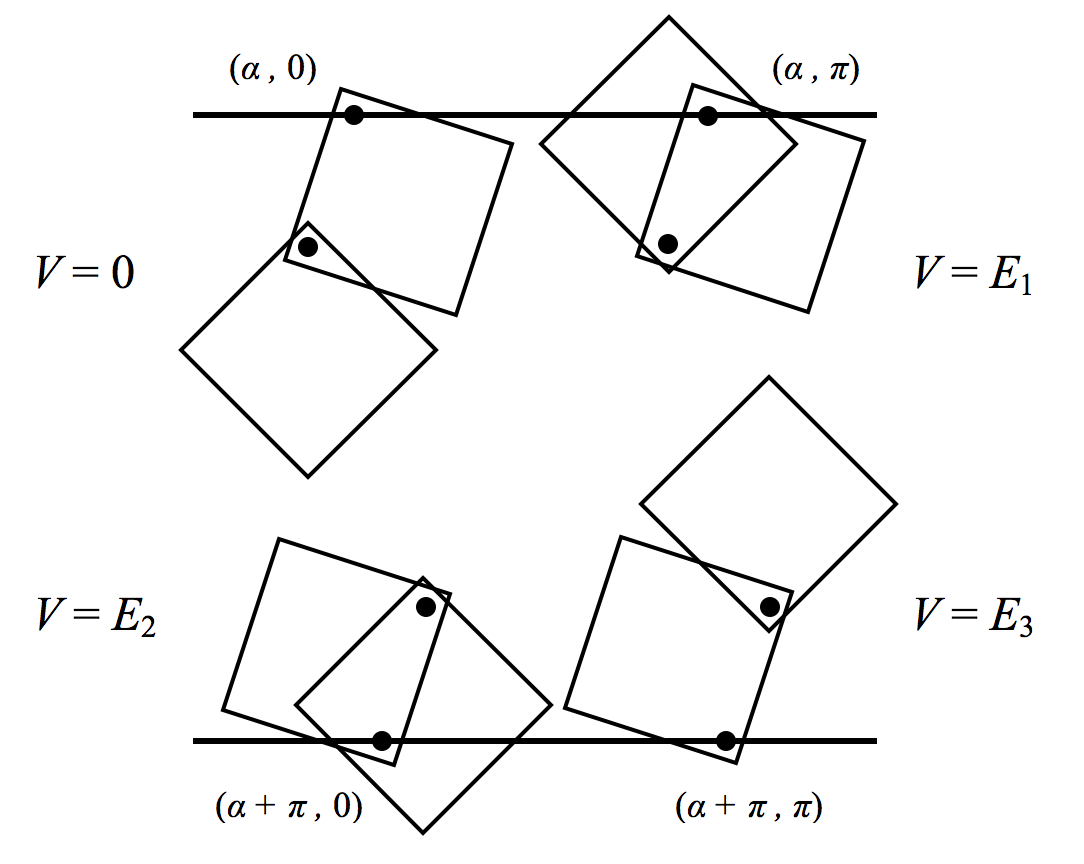}}
\caption{\label{fig:equilibs}The equilibrium configurations of the double 
square pendulum, with the stable equilibrium at upper left. The axles $P$ and
$Q$ are indicated by black circles, and the angles $(\theta_1,\theta_2)$ and
potential energies of the configurations are also shown.}
\end{center}
\end{figure}

Equilibrium configurations of the pendulum occur when 
$V =V(\theta_1,\theta_2 )$ is
stationary with respect to $\theta_1$ and $\theta_2$.
The four equilibrium configurations are
\begin{equation}
(\theta_1, \theta_2) = (\alpha, 0), \quad (\alpha, \pi), \quad 
(\alpha + \pi, 0), \quad 
\textrm{and} \quad (\alpha + \pi, \pi);
\end{equation}
but $(\theta_1, \theta_2) = (\alpha, 0)$ is the only stable equilibrium. 
The equilibrium configurations are illustrated in Fig.~\ref{fig:equilibs} with the 
stable equilibrium at the upper left. The angle $\alpha$, defined by 
Eq.~(\ref{eqn:alpha}), is the angle that the center of mass of the inner 
plate makes with the negative $y$ axis in stable equilibrium.

It is convenient to choose $V_0$ so that the potential energy of the pendulum 
is zero in stable equilibrium. In this case the potential energies of the 
pendulum in the three unstable equilibrium configurations are $V=E_i$ 
(with $i=1,\,2,\,3$), where
\begin{subequations}\label{eqn:Erange}
\begin{eqnarray}
E_1 & = & \sqrt{2} m_2 g \ell,\\
E_2 & = & E_1\,\mathrm{cosec}\,\alpha, \\
E_3 & = & E_1+E_2.
\end{eqnarray}
\end{subequations}
The configurations shown in Fig.~\ref{fig:equilibs} are labeled by these 
energies.

It is convenient to remove the dependence of the equilibrium 
coordinates on $\alpha$ by introducing the change of 
coordinates
\begin{equation}\label{eqn:coord_change}
\varphi_1 = \theta_1 - \alpha \quad \textrm{and} \quad \varphi_2 = \theta_2.
\end{equation}
With these choices the potential energy of the pendulum may be written as
\begin{equation}
\label{eqn:V}
V = \frac{1}{2} \left[(1 - \cos \varphi_1) E_1 + 
(1 - \cos \varphi_2)E_2\right].
\end{equation}
The simple double pendulum has analogous equilibrium configurations,
and its potential energy may also be expressed in the form of 
Eq.~(\ref{eqn:V}).\cite{korsch99}

In terms of the coordinates in Eq.~(\ref{eqn:coord_change}) the 
kinetic energy of the pendulum in Eq.~(\ref{eqn:T_gen}) may be written as
\begin{equation}\label{eqn:T}
T = m_2(k_2 {\dot{\varphi}_1}^2 + 2 k_3 \dot{\varphi}_1 \dot{\varphi}_2 
\sin \beta + k_4 \dot{\varphi}_2^2),
\end{equation}
where 
\begin{align}
k_2 &= \frac{1}{12}\left[(m_1/m_2) L^2 + 3 (m_1/m_2 + 2) \ell^2\right], \\
k_3 &= \sqrt{2}\ell^2/4, 
\qquad
k_4 = \frac{1}{12}(L^2 + 3 \ell^2), 
\end{align}
and 
\begin{equation}
\beta = \pi/4 + \alpha + \varphi_1 - \varphi_2.
\end{equation}

We apply the Euler-Lagrange equations\cite{landau76,goldstein80}
to the Lagrangian $\mathcal{L} = T - V$ given by Eqs.~(\ref{eqn:V}) 
and (\ref{eqn:T}) and obtain the equations of motion:
\begin{subequations}
\label{eqn:eom}
\begin{eqnarray}
k_2 \ddot{\varphi}_1 + k_3(\ddot{\varphi}_2\sin\beta - \dot{\varphi}_2^2 
\cos\beta) + k_1 \sin \varphi_1 &=& 0, \\
k_4 \ddot{\varphi_2} + k_3(\ddot{\varphi}_1\sin\beta + \dot{\varphi}_1^2
\cos\beta) + k_5 \sin\varphi_2 &=& 0,
\end{eqnarray}
\end{subequations}
where 
\begin{equation}
k_5 = \sqrt{2}g \ell/4. 
\end{equation}

\section{General Features of the Motion \label{sec:bchar}}

The coupled, nonlinear equations of motion in \eqref{eqn:eom} are not 
amenable to 
analytic solution, and it is necessary to
solve these equations numerically to investigate the motion.
Some general statements can be made about the behavior for 
a given energy, and about the nature of the motion at small and large 
energies.

The motion of the pendulum depends on its total energy $E = T + V$.
The magnitude of the energy in relation to the three energies $E_1$, $E_2$, 
and $E_3$ specified by Eq.~(\ref{eqn:Erange}) determines whether the 
plates of the pendulum can perform complete rotations.
For $E \leq E_1$ each of the plates may oscillate about stable
equilibrium, but there is insufficient energy for either plate to
perform a complete rotation. For $E_1 < E \leq E_2$ the energy of the 
pendulum is sufficient
to allow complete rotation of the outer plate about the axle at $Q$ (see
Fig.~\ref{fig:model}), but rotational motion of the inner plate is still
prohibited. One or other of the plates may rotate for $E_2 < E \leq E_3$,
and simultaneous rotation becomes possible when $E > E_3$. The simple
double pendulum has analogous behavior.\cite{korsch99}

\subsection{Motion at low energy\label{sec:bchar:normal}}

The nonlinear terms in the equations of motion have negligible influence
when the total energy is small, in which case the pendulum oscillates 
with small amplitude about stable equilibrium. In this regime
the equations may be simplified by using small-angle approximations and 
dropping nonlinear terms,\cite{hand98,kibble05,gregory06} leading to 
the linear equations:
\begin{subequations}\label{eqn:eom_lin}
\begin{align}
k_2 \ddot{\varphi}_1 + k_6 \ddot{\varphi}_2 + k_1 \varphi_1 &= 0, \\
k_4 \ddot{\varphi}_2 + k_6 \ddot{\varphi}_1 + k_5 \varphi_2 &= 0,
\end{align}
\end{subequations}
where 
\begin{equation}
k_6=k_3(\cos\alpha+\sin\alpha)/\sqrt{2}.
\end{equation}
These equations have the general form expected for coupled linear 
oscillators.\cite{main84}

Normal modes of oscillation are motions of the pendulum in which the 
coordinates $\varphi_1$ and $\varphi_2$ vary harmonically in 
time with the same frequency and phase, but not necessarily with 
the same amplitude.\cite{gregory06} 
The substitution of harmonic solutions into the linearized 
equations of motion \eqref{eqn:eom_lin} leads to the identification
of two normal frequencies $\omega_{+}$ and $\omega_{-}$, 
corresponding to fast and slow modes of oscillation:
\begin{equation}\label{eqn:freqs}
\omega_{\pm}^2 = \frac{k_1k_4+k_2k_5\pm \sqrt{(k_1k_4-k_2k_5)^2+2k_1k_5k_6^2}}{2a},
\end{equation}
where $a = k_2 k_4 - k_6^2$. The amplitude 
factors $A_1$ and $A_2$ for the harmonic motions of the two
coordinates are related by
\begin{equation}
\Big(\frac{A_1}{A_2} \Big)_{\pm} = \frac{-k_5 k_6} {a\omega_{\pm}^2-k_1k_4}.
\end{equation}
For the slow mode $(A_1/A_2)_{+}>0$, so that the plates oscillate 
in the 
same direction; for the fast mode $(A_1/A_2)_{-}<0$, and 
the plates
oscillate in opposite directions.

A simple case considered in Secs.~\ref{sec:num} and \ref{sec:poinc}
occurs when $m_1 = m_2$ and 
$\ell = L$; that is, the plates have equal mass and the axles are located
at the corners of the plates. In that case the normal frequencies are
$\omega_{+} \approx 1.66 (g/L)^{1/2}$ and 
$\omega_{-} \approx 0.782 (g/L)^{1/2}$, and the amplitudes 
are $A_1 \approx -0.613 A_2$ for the fast mode and $A_1 \approx 0.730 A_2$ 
for the slow mode.

The general motion at low energy may be expressed as linear combinations 
of the normal modes,\cite{main84} in which case the motion is no longer 
periodic, but is quasi-periodic. The motion never quite repeats itself
for general initial conditions. The simple double 
pendulum has analogous behavior at low energy.\cite{korsch99}

\subsection{Motion at high energies\label{sec:bchar:high}}

At high energy the pendulum behaves like a simple rotor, with the 
system rotating rapidly in a stretched configuration 
($\theta_1\approx \pi/4$, $\theta_2\approx 0$). In this case
the kinetic energy terms in the Lagrangian dominate the
potential energy terms and may be described by setting $g=0$ in
the equations of motion. The total angular momentum is conserved, 
because in the absence of gravity, there is no torque on the pendulum. 
The resulting motion of the system is regular 
(non-chaotic), 
because a system with two degrees of freedom and two constraints 
(conservation of total energy and total angular momentum) cannot exhibit 
chaos.\cite{richter84} It follows, for example, that the double square
pendulum would not exhibit chaos if installed on the space station. 
The simple double pendulum has analogous behavior.\cite{korsch99}

\section{Numerical Methods \label{sec:num}}

To investigate the detailed dynamics of the pendulum the equations of 
motion are solved numerically by using the fourth-order 
Runge-Kutta method.\cite{press92} 
The accuracy of the integration may be checked by evaluating the energy 
of the pendulum at each integration step. If there is a discrepancy between 
the calculated energy and the initial energy of the pendulum, the integration 
step may be modified accordingly.

The equations of motion of the pendulum may be written in dimensionless form by
introducing
\begin{equation}
\overline{t} = \sqrt{\frac{g}{L}} t,\quad \overline{L} = \frac{\ell}{L},
\quad \text{and} \quad \overline{m} = \frac{m_1}{m_2}.
\end{equation}
An appropriate dimensionless energy is 
\begin{equation}
\overline{E} = \frac{E}{\frac{1}{12}m_2 g L}. 
\end{equation}

In Sec.~\ref{sec:poinc} the system is simplified by considering equal
mass plates and by locating the axles at the corners of the plates. These
choices imply parameter values
\begin{equation}
\overline{L} = 1 \quad \text{and} \quad \overline{m} = 1.
\end{equation}
With these choices, the energies of the pendulum corresponding
to Eq.~\eqref{eqn:Erange} are
$\overline{E}_1 \approx 16.97$, 
$\overline{E}_2 \approx 37.95$, 
and $\overline{E}_3 \approx 54.92$. In Sec.~\ref{sec:poinc} we also
use dimensionless variables and drop the
bars for simplicity.

\section{General Dynamics and Poincar\'{e} Sections\label{sec:poinc}}

The general dynamics of the pendulum may be investigated by analyzing the
phase space for increasing values of the total energy. The phase space of
the pendulum is three-dimensional. (There are four coordinates, that is,
$\varphi_1 $, $\varphi_2$, $\dot{\varphi}_1$, and 
$\dot{\varphi }_2 $, but one of these may be eliminated because 
energy is conserved.) The
phase space can be examined by considering the two-dimensional
Poincar\'{e} section,\cite{drazin92,korsch99,hilborn00} defined by
selecting one of the phase elements and plotting the values of others
every time the selected element has a certain value. For a given choice 
of initial conditions the Poincar\'{e} section shows points representing
the intersection of an orbit in phase space with a plane
in the phase space. Periodic orbits produce a finite set of points in
the Poincar\'{e} section, quasi-periodic orbits produce a continuous
curve, and chaotic orbits result in a scattering of points within
an energetically accessible region.\cite{drazin92,korsch99}

\subsection{Defining a Poincar\'{e} section\label{sec:poinc:def}}

We use conservation of energy to eliminate $\dot{\varphi}_2$
and choose the Poincar\'{e} plane to be $\varphi_2 = 0$. A point 
$(\varphi_1, \dot{\varphi}_1)$ is recorded in the phase space of the
inner plate whenever the outer plate passes through the vertical position 
$\varphi_2 = \theta_2 = 0$. When
this condition occurs, the outer plate may have positive or negative 
momentum. To
ensure a unique definition of the Poincar\'{e} section a point is 
recorded
if the outer plate has positive momentum; that is, a point is recorded 
in the
phase space of the inner plate whenever
\begin{equation}
\varphi_2 = 0 \qquad \textrm{and} \qquad {p}_{2} > 0,
\end{equation}
where
$p_{2} = \partial {\mathcal{L}}/\partial 
\dot{\varphi}_2 = 6\sqrt{2}\dot{\varphi}_1 \sin\beta + 8\dot{\varphi}_2$
is the generalized momentum corresponding to $\varphi_2$. In the following
we present Poincar\'{e} sections as plots of $\dot{\theta}_1$ versus 
$\theta_1$, with the angles in degrees.

\subsection{Results\label{sec:poinc:results}}

\begin{figure}[!ht]
\begin{center}
\scalebox{0.5}{\includegraphics{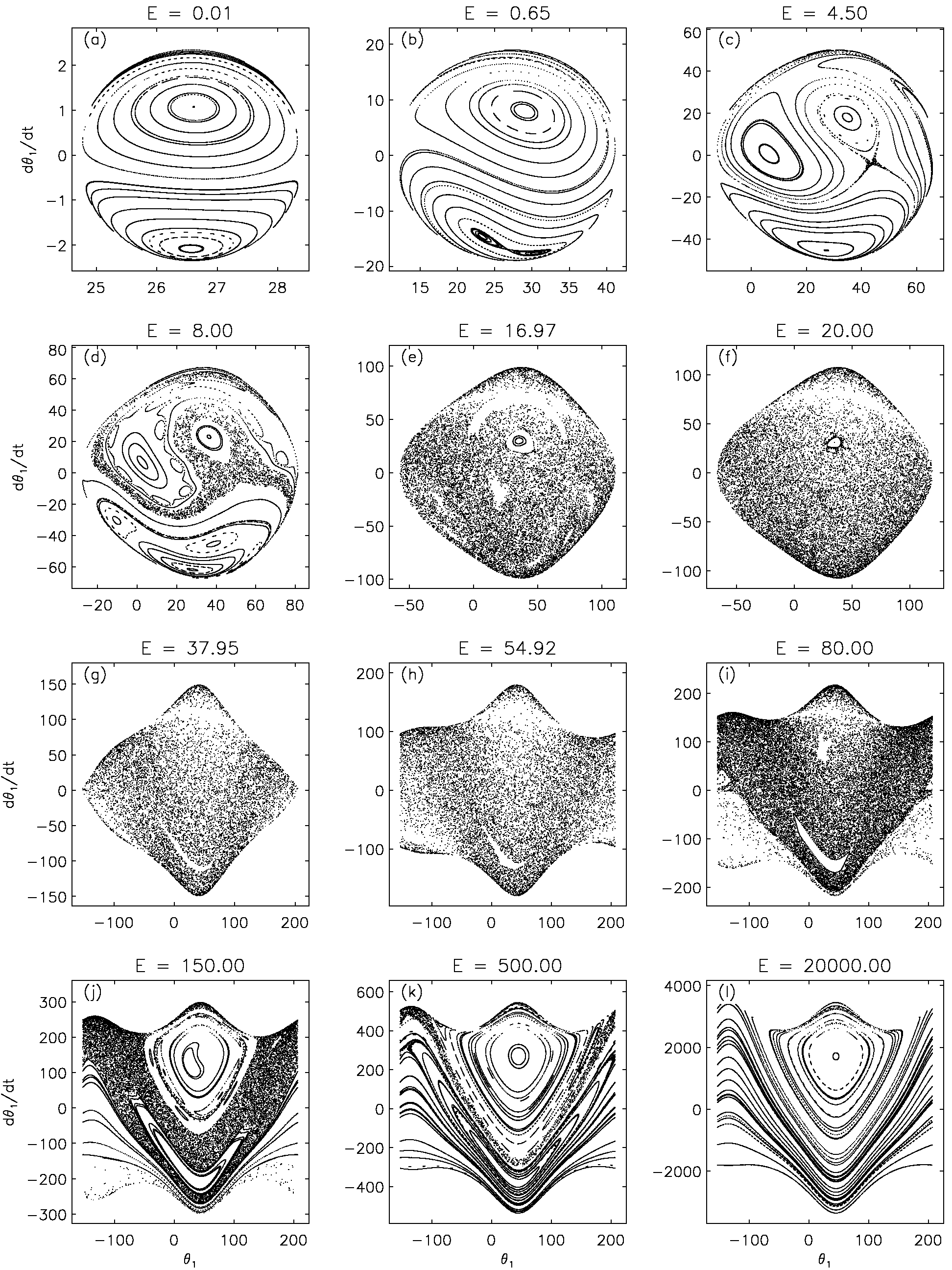}}
\caption{\label{fig:poinc}Poincar\'{e} sections of the 
double square pendulum for increasing values of the dimensionless total 
energy $E$. The panels show: (a) $E=0.01$; (b) $E=0.65$; (c)
$E=4.50$; (d) $E=8$; (e) $E=E_1=16.97$; (f) $E=20$; (g) $E=E_2=37.95$;
(h) $E=E_3=54.92$; (i) $E=80$; (j) $E=150$; (k) $E=500$; and
(l) $E=20000$.}
\end{center}
\end{figure}

Figure~\ref{fig:poinc} shows the results of constructing Poincar\'{e}
sections for values of the total
(dimensionless) energy ranging from $E=0.01$ to 
$E=2\times 10^4$.\cite{wheatland08b}
Each section was
constructed by numerically solving Eq.~(\ref{eqn:eom_lin}) at a given
energy for many different initial conditions; 40--60 initial conditions
were used for the cases shown in Fig.~\ref{fig:poinc}, with initial conditions
chosen to provide a good coverage of the energetically accessible
region in the plane.

For energy $E = 0.01$ [Fig.~\ref{fig:poinc}(a)] the Poincar\'{e}
section is covered by two regions of stable elliptical orbits around two
fixed points located on the line $\theta_1 = \alpha \approx 26.6^\circ$,
and this entire section has approximate reflection symmetry about this line. 
The motion
of the pendulum is regular (that is, periodic and quasi-periodic) at low
values of its total energy. The fixed points correspond to the two strictly
periodic normal modes identified in Sec.~\ref{sec:bchar:normal}, with
the upper fixed point corresponding to the co-oscillating slow mode, and
the lower fixed point corresponding to the counter-oscillating fast mode.
The behavior observed in Fig.~\ref{fig:poinc}(a) corresponds to the
solution of the linearized equations of motion in Eq.~(\ref{eqn:eom_lin}),
and is strictly regular. The horizontal and vertical extent of the section
is very small (a few degrees) due to the small energy.

Figure~\ref{fig:poinc}(b) shows the Poincar\'{e} section 
for $E = 0.65$. The section has lost the reflection 
symmetry about $\theta_1=\alpha\approx 26.6^\circ$ observed in 
Fig.~\ref{fig:poinc}(a). This loss of symmetry is related to the 
non-symmetrical geometry of the pendulum
(the center of mass of the inner plate is offset to the right).
The section is much larger, spanning almost $30^\circ$ in $\theta_1$,
and almost $40^\circ$ in $\dot{\theta}_1$ (recall that time is
dimensionless), due to the larger energy.
A striking feature of this section is that the lower fixed point 
has undergone a period-doubling 
bifurcation, resulting in a pair of stability islands separated by a 
separatrix within the lower stability region. A trajectory 
producing an orbit in either of these stability islands produces a 
corresponding orbit in the other due to the period $2$ nature of the 
orbit. 

The first signs of chaotic behavior appear at $E \approx 4$--4.5
[see Fig.~\ref{fig:poinc}(c) for $E=4.5$] in the form of a scattering
of points around the hyperbolic point that sits between the four stability 
regions. The chaotic region first appears along the trajectory containing 
the hyperbolic point. As the energy is increased from $E = 4.5$ to 
$E = 8$ [Fig.~\ref{fig:poinc}(d)] the size of the chaotic region
increases with the loss of regular orbits, particularly in the stability
regions located at the upper left and upper right of the Poincar\'{e}
section. The lower stability region remains unaffected.

The growth of chaos with energy is demonstrated by the next three 
Poincar\'{e} sections, corresponding to $E = E_1=16.97$ 
[Fig.~\ref{fig:poinc}(e)], $E = 20$ [Fig.~\ref{fig:poinc}(f)], and
$E=E_2=37.95$ [Fig.~\ref{fig:poinc}(g)]. Figure~\ref{fig:poinc}(e) 
corresponds to the pendulum
having sufficient energy for the outer plate to rotate. It is striking
that the Poincar\'{e} section has few regions with 
stable orbits, so that the pendulum is chaotic even at this modest
energy. At $E = 20$ [Fig.~\ref{fig:poinc}(f)] most of the 
invariant orbits in the upper stability region have been lost. Notably
the upper fixed point persists, and is surrounded by small regions
of stability. The upper fixed point corresponds to a large amplitude
analog of the slow normal mode.
Global chaos is achieved at energy $E \approx 25$ (not shown) where a 
single trajectory covers the entire Poincar\'{e} section and the system 
is completely ergodic. Note that this energy is less than $E_2=37.95$,
the energy required for the inner plate to rotate. Hence completely
chaotic behavior is achieved even without the rotation of the inner plate.
Global chaos remains up to 
$E \approx 37$ at which energy a small stability island 
appears in the lower part of the Poincar\'{e} section and grows with 
energy. Figure~\ref{fig:poinc}(g) shows $E=E_2=37.95$.
The Poincar\'{e} section has grown to a width of $360^\circ$ in 
$\theta_1$. 

A second stability island 
appears near the upper part of the Poincar\'{e} section at 
$E\approx 75$. This stability island corresponds to quasi-periodic
rotational motion of the entire pendulum, and is an very large amplitude
analog of the slow normal mode. The two stability islands, 
plus a third stability island located on the left and right hand edges, 
are shown in 
the Poincar\'{e} section at $E = 80$ [Fig.~\ref{fig:poinc}(i)].

The sizes of the stability regions increase with energy, as shown by the
cases $E = 150$ [Fig.~\ref{fig:poinc}(j)] and $E = 500$ 
[Fig.~\ref{fig:poinc}(k)]. 
The motion of the pendulum becomes regular at very high values 
of $E$, as shown in Fig.~\ref{fig:poinc}(l), corresponding to 
$E = 2\times 10^4$. The appearance of regular behavior at high
energies was discussed in Sec.~\ref{sec:bchar:high}, and explained in 
terms of the total angular momentum being conserved during
the motion, in addition to the total energy. The section also becomes 
approximately reflection symmetric at this energy, about the line
$\theta_1=45^\circ$.

\subsection{Comparison with simple double pendulum\label{sec:poinc:simple}}

The geometry of the Poincar\'{e}
sections presented in Sec.~\ref{sec:poinc:results} show some
differences from those of the simple double pendulum and double bar 
pendulum.\cite{korsch99, ohlhoff00, richter84,stachowiak06} To better
illustrate the differences we have numerically solved the equations
of motion for the simple double pendulum, and constructed Poincar\'{e}
sections using the definition given in Sec.~\ref{sec:poinc:def}. 
For simplicity we
consider a simple double pendulum with equal masses $M$, connected
by massless rods with equal length $h$. The energy of the system is given in terms 
of $\frac{1}{12}M g h$, and time is in terms
of $\sqrt{h/g}$. The dimensionless energies corresponding to
threshold values for complete rotation of the inner mass, outer mass,
and both masses are $E_{1,\rm s}=24$, $E_{2,\rm s}=48$, and 
$E_{3,\rm s}=72$ respectively,
where the subscript $s$ denotes the simple double 
pendulum. The energies are equally spaced due to the symmetry of 
the pendulum.

\begin{figure}[!ht]
\begin{center}
\scalebox{0.5}{\includegraphics{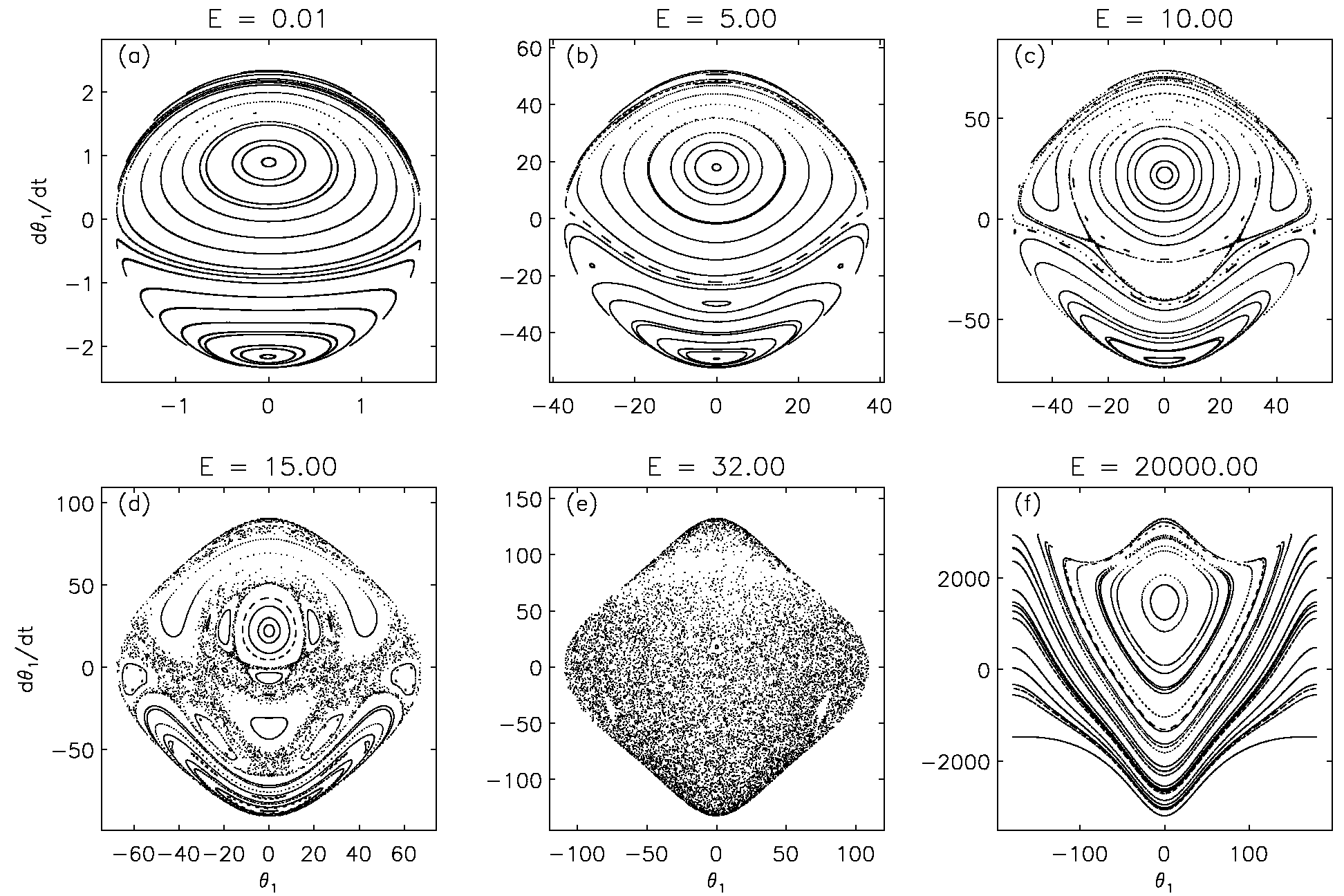}}
\caption{\label{fig:poinc_simple}Poincar\'{e} sections of the 
simple double pendulum for increasing values of $E$. The panels
show: (a) $E=0.01$; (b) $E=5$; (c) $E=10$; (d) $E=15$; (e) $E=32$;
and (f) $E=20000$.
}
\end{center}
\end{figure}

Figure~\ref{fig:poinc_simple} illustrates a sequence of Poincar\'{e}
sections for the simple double pendulum, which may be considered
to be comparable to certain sections shown in Fig.~\ref{fig:poinc}
for the double square pendulum, as described in the following.

At $E = 0.01$ [Fig.~\ref{fig:poinc_simple}(a)] the Poincar\'{e}
section is very similar to that obtained for the double square 
pendulum at the same energy [Fig.~\ref{fig:poinc}(a)], except that 
the section for the simple double pendulum 
is symmetric about $\theta_1=0$ rather than $\theta_1=\alpha$, 
and the orbits in the section
have exact reflection symmetry about this line (rather than approximate
symmetry). All of the Poincar\'{e}
sections for the simple double pendulum have strict reflection symmetry 
about $\theta_1=0$.
In common with the double square pendulum the section at the lowest
energy consists of regular
orbits about two fixed points corresponding to two normal modes.

At somewhat higher energies [Fig.~\ref{fig:poinc_simple}(b)] the orbits 
in the Poincar\'{e} section distort. Figure~\ref{fig:poinc_simple}(b)
is comparable to Fig.~\ref{fig:poinc}(b) with two notable
differences. The Poincar\'{e} section for the double square pendulum 
exhibits period-doubling of the lower fixed point, which has no 
counterpart for the simple double pendulum (or for the double bar 
pendulum),\cite{korsch99, ohlhoff00, richter84, stachowiak06} and
may be related to the non-symmetrical geometry of the double square
pendulum. The other obvious difference is the loss of symmetry in 
the section for the double square pendulum.

Figure~\ref{fig:poinc_simple}(c) illustrates the first appearance of chaos
in the simple double pendulum; this section may be considered to be
comparable to Fig.~\ref{fig:poinc}(c). Note that the 
double square pendulum first becomes chaotic at an energy substantially 
lower than that for the simple double pendulum in comparison to the 
respective threshold energies required for complete rotation of the 
masses. For the double square pendulum chaos appears at an energy 
$E_{\rm c}\approx 4$, a fraction $E_{\rm c}/E_1\approx 0.24$ of the energy required for
the outer plate to rotate, and a fraction $E_{\rm c}/E_3\approx 0.07$ of the
energy required for both plates to rotate. In comparison, chaos first
appears in the simple double pendulum at $E_{\rm c,s}\approx 10$, 
which
corresponds to $E_{\rm c,s}/E_{1,\rm s}\approx 0.42$ and 
$E_{\rm c,s}/E_{3,\rm s}\approx 0.14$ of the energy required for rotation 
of the 
outer mass and of both masses, respectively. This difference may be due 
to the 
additional complexity introduced into the motion by the geometrical
asymmetry. Figures~\ref{fig:poinc_simple}(c) ($E=10$) 
and \ref{fig:poinc_simple}(d) ($E=15$) illustrate
the development of chaos, and are comparable to Figs.~\ref{fig:poinc}(c)
and \ref{fig:poinc}(d). 

Figure~\ref{fig:poinc_simple}(e) shows the Poincar\'{e} section for 
$E=32$; this section is comparable with Fig.~\ref{fig:poinc}(f). 
In both cases the pendula have sufficient energy for the lower mass, but 
not the upper mass, to rotate. Both Poincar\'{e} sections are chaotic, 
apart from stable regions around the upper fixed point, which corresponds to 
a large amplitude slow mode.

The geometry of the Poincar\'{e} section at very high energy 
($E=2\times 10^4$) shown in Fig.~\ref{fig:poinc_simple}(f) is very similar 
to that of the double square pendulum at the same energy
[Fig.~\ref{fig:poinc}(l)]. This similarity is expected because at high 
energies the different pendula perform rotational motion in a 
stretched configuration, and the difference in geometry is of little
importance. The section for the simple double pendulum is exactly
symmetrical about $\theta_1=0$, whereas the section for the double square
pendulum is approximately symmetrical about $\theta_1=45^\circ$.

\section{Qualitative comparison with a real double square pendulum \label{sec:device}}

A detailed comparison with a real double square pendulum
requires an experimental investigation and solution of the equations 
of motion including the more complicated distributions of masses in 
that double square pendulum, which is beyond the scope of this article. In
this section we make some brief qualitative comparisons between the 
results of the model and the behavior of the double square pendulum.

As mentioned in Sec.~I, the real double square pendulum exhibits regular 
behavior at high and low energies and irregular behavior at intermediate
energies. One way to demonstrate the
range of behavior of a real double square pendulum is to set it into motion with high
energy, and to watch the change in behavior as the double square pendulum slowly loses 
energy due to friction. 

It is straightforward to demonstrate each of the normal modes in a 
real double square pendulum by turning the central wheel back and forth with small
amplitude and with the appropriate frequencies. By timing multiple 
oscillations, the periods of the normal modes were measured to be 
$T_{+}\approx 0.70\pm 0.01\,{\rm s}$ 
and
$T_{-} \approx 1.23\pm 0.01\,{\rm s}$. For equal mass plates with
axles at the corners of the plates ($\ell=L$), and
$L=0.28\,{\rm m}$ and $g=9.81\,{\rm m}\,{\rm s}^{-2}$, Eq.~(\ref{eqn:freqs})
predicts $T_{+}\approx 0.64\,{\rm s}$ and $T_{-}\approx 1.36\,{\rm s}$. 
If we include an offset of the axles ($\ell=0.1 L$), the result is 
$T_{+}\approx 0.64\,{\rm s}$ and $T_{-}\approx 1.25\,{\rm s}$. 
The slow mode period predicted by the model is approximately correct, but the predicted fast 
mode period is too small by about 10\%. These results illustrate the 
relative accuracy of the simple model.

We also demonstrated the 
appearance of chaos in the double square pendulum. By turning the 
central wheel through $180^\circ$ from stable equilibrium, the 
double square pendulum may be put into the unstable equilibrium 
configuration corresponding to energy $E_2$ in the 
model [the lower left configuration in Fig.~3]. 
The double square pendulum may be held at rest in this position 
and then released. The model predicts that the double square 
pendulum is almost completely chaotic at this energy, as shown 
by the Poincar\'{e} section in Fig.~4(g), and in particular the 
model is chaotic with these initial conditions. The motion of the 
double square pendulum was video taped several times after release 
from this initial configuration. Comparison of the corresponding 
movie frames (with the correspondence determined by the initial 
motion) shows that the motion is the same for a few rotations 
and oscillations of the plates, and then rapidly becomes different, 
providing a striking demonstration of the sensitivity to initial 
conditions characteristic of chaos.\cite{wheatland08c}

\section*{Acknowledgments}

The authors thank Dr Georg Gottwald, Professor Dick Collins,
and Dr Alex Judge for advice, comments, and assistance. The paper
has benefited from the work of two anonymous reviewers.

\section*{Figure captions}

\begin{thebibliography}{50}

\bibitem{richter84} P. H. Richter and H. -J. Scholz, 
``Chaos in classical mechanics: The double pendulum,'' 
in \textit{Stochastic Phenomena and Chaotic Behavior in 
Complex Systems}, edited by P. Schuster (Springer, Berlin, 
Heidelberg, 1984), pp. 86--97.

\bibitem{korsch99} H. J. Korsch and H. -J. Jodl, 
\textit{Chaos: A Program Collection for the PC} 
(Springer-Verlag, Berlin, 1999), 2nd ed.

\bibitem{stachowiak06} T. Stachowiak and T. Okad, 
``A numerical analysis of chaos in the double pendulum,'' 
Chaos, Solitons and Fractals \textbf{29}, 417--422 (2006).

\bibitem{landau76} L. D. Landau and E. M. Lifshitz, 
\textit{Mechanics} (Elsevier, Amsterdam, 1976), 3rd ed.

\bibitem{goldstein80} H. Goldstein, 
\textit{Classical Mechanics} 
(Addison-Wesley, Reading MA, 1980), 2nd ed.

\bibitem{kibble05} T. W. B. Kibble and F. H. Berkshire, 
\textit{Classical Mechanics} (Imperial College Press, 
London, 2005), 5th ed.

\bibitem{gregory06} R. D. Gregory, \textit{Classical 
Mechanics} (Cambridge University Press, Cambridge, 2006).

\bibitem{newton89} P. K. Newton, 
``Escape from Kolmogorov-Arnold-Moser regions and breakdown 
of uniform rotation,'' Phys. Rev. A \textbf{40}, 3254--3264 
(1989).

\bibitem{levien93} R. B. Levien and S. M. Tan, 
``Double pendulum: An experiment in chaos,'' 
Am. J. Phys. \textbf{61}, 1038--1044 (1993).

\bibitem{shinbrot92} T. Shinbrot, C. Grebogi, J. Wisdom, 
and J. A. Yorke, ``Chaos in a double pendulum,'' 
Am. J. Phys. \textbf{60}, 491--499 (1992).

\bibitem{zhou96} Z. Zhou and C. Whiteman, 
``Motions of double pendulum,'' 
Nonlinear Analysis, Theory, Method and Applications 
\textbf{26} (7), 1177--1191 (1996).

\bibitem{ohlhoff00} A. Ohlhoff and P. H. Richter, 
``Forces in the double pendulum,'' 
Z. Angew. Math. Mech. \textbf{80} (8), 517--534 (2000).

\bibitem{chaoticpendulums} Available at 
\url{<http://www.chaoticpendulums.com/>}.

\bibitem{dullin94} H. R. Dullin, 
``Melkinov's method applied to the double pendulum,'' 
Z. Phys. B \textbf{93}, 521--528 (1994).

\bibitem{wheatland08a} Movies showing the double 
square pendulum in action are available at 
\url{<www.physics.usyd.edu.au/~wheat/sdpend/>}.

\bibitem{hand98} L. N. Hand and J. D. Finch 
\textit{Analytical Mechanics} 
(Cambridge University Press, Cambridge, 1998).

\bibitem{main84} I. G. Main, 
\textit{Vibrations and Waves in Physics} 
(Cambridge University Press, Cambridge, 1984), 2nd ed.

\bibitem{press92} W. H. Press, B. P. Flannery, 
S. A. Teukolsky and W. T. Vetterling, 
\textit{Numerical Recipes in C} 
(Cambridge University Press, New York, 1992), 2nd ed.

\bibitem{drazin92} P. G. Drazin, 
\textit{Nonlinear Systems} 
(Cambridge University Press, Cambridge, 1992).

\bibitem{hilborn00} R. C. Hilborn, 
\textit{Chaos and Nonlinear Dynamics: 
An Introduction for Scientists and Engineers} 
(Oxford University Press, New York, 2000), 2nd ed.

\bibitem{wheatland08b} Animations of the numerical 
solution of the equations of motion for different 
energies corresponding to some of the choices in F
ig.~\ref{fig:poinc} are available at 
\url{<www.physics.usyd.edu.au/~wheat/sdpend/>}.

\bibitem{tel06} T. T\'{e}l and M. Gruiz, 
\textit{Chaotic Dynamics: 
An Introduction Based on Classical Mechanics} 
(Cambridge University Press, Cambridge, 2006).

\bibitem{wheatland08c} A movie showing nearly aligned 
frames for three releases of the pendulum is available 
at \url{<www.physics.usyd.edu.au/~wheat/sdpend/>}.

\end{thebibliography}
\end{document}